# The good, the bad and the user in soft matter simulations

Jirasak Wong-ekkabut[1,*] and Mikko Karttunen[2,*]

[1]Department of Physics, Faculty of Science, Kasetsart University, Bangkok, 10900, Thailand

[2]Department of Mathematics and Computer Science & Institute for Complex Molecular Systems, Eindhoven University of Technology, MetaForum, 5600 MB Eindhoven, the Netherlands

**\*Corresponding Authors**

E-mail address J.W.: jirasak.w@ku.ac.th, and M.K.: mkarttu@tue.nl

## Contents

## Abstract

Molecular dynamics (MD) simulations have become popular in materials science, biochemistry, biophysics and several other fields. Improvements in computational resources, in quality of force field parameters and algorithms have yielded significant improvements in performance and reliability. On the other hand, no method of research is error free. In this review, we discuss a few examples of errors and artifacts due to various sources and discuss how to avoid them. Besides bringing attention to artifacts and proper practices in simulations, we also aim to provide the reader with a starting point to explore these issues further. In particular, we hope that the discussion encourages researchers to check software, parameters, protocols and, most importantly, their own practices in order to minimize the possibility of errors. The focus here is on practical issues.

# 1. Introduction

Computer simulations have become an extremely powerful tool and have firmly established their role as the third paradigm of research alongside theory and experiments as demonstrated by the 2013 Nobel Prize in Chemistry to Martin Karplus, Michael Levitt and Arieh Warshel *"for the development of multiscale models for complex chemical systems"*. Simulations have predictive power. For example, they have predicted that proteins are dynamic entities, that their dynamics is essential for their biological functions[1], that highly hydrophobic carbon nanotubes can conduct water[2], and elucidated the functioning principles of aquaporins[3, 4]. In our own computational research, we have, e.g., resolved the decades old problem of lipid diffusion modes[5]. Recent reviews discussing the progress and success of simulations are provided, for example, by Schulten et al.[6], Dror et al.[7], and Karplus and McCammon[8]. In this review, we focus on classical level simulations and will not address *ab initio* simulations. A recent review discussing various aspects of *ab initio* simulations is provided by Kirchner et al.[9]. Although lattice Boltzmann simulations can contain embedded particles at the classical MD level, we will not discuss them or hybrid simulations. Instead, we refer the reader to Silva and Semiao[10] who discuss the lattice structure and truncation errors and Ollila et al.[11] who provide a method to properly include thermal fluctuations and thermostatting in lattice Boltzmann.

In addition to being predictive, a less appreciated important fact about simulations is they can show the kind of disturbances experimental probes, for example fluorescent probes or local heating by laser, can cause[12, 13]. Another under-appreciated fact is that several experimental methods need at least a semi-computational molecular model for interpretation of the experimental data. This is particularly the case with NMR (Nuclear Magnetic Resonance) and various scattering methods. Typically, such models are not true simulations but rather just models for interpretation of experimental data. Proper simulations, as already stated by Allen and Tildesley in their classic book "Computer Simulations of Liquids"[14], have the ability to be predictive and to provide exact results for a given model within algorithmic and numerical accuracy. This is indeed remarkable! No other method has the latter ability: No mean-field approximations are necessary and the disturbances of external probes are absent. Yet at the same time, all the positions and velocities of all atoms are known and can be post-analyzed in any way desirable. Relations between experiments, computation and theory are illustrated in Figure 1.

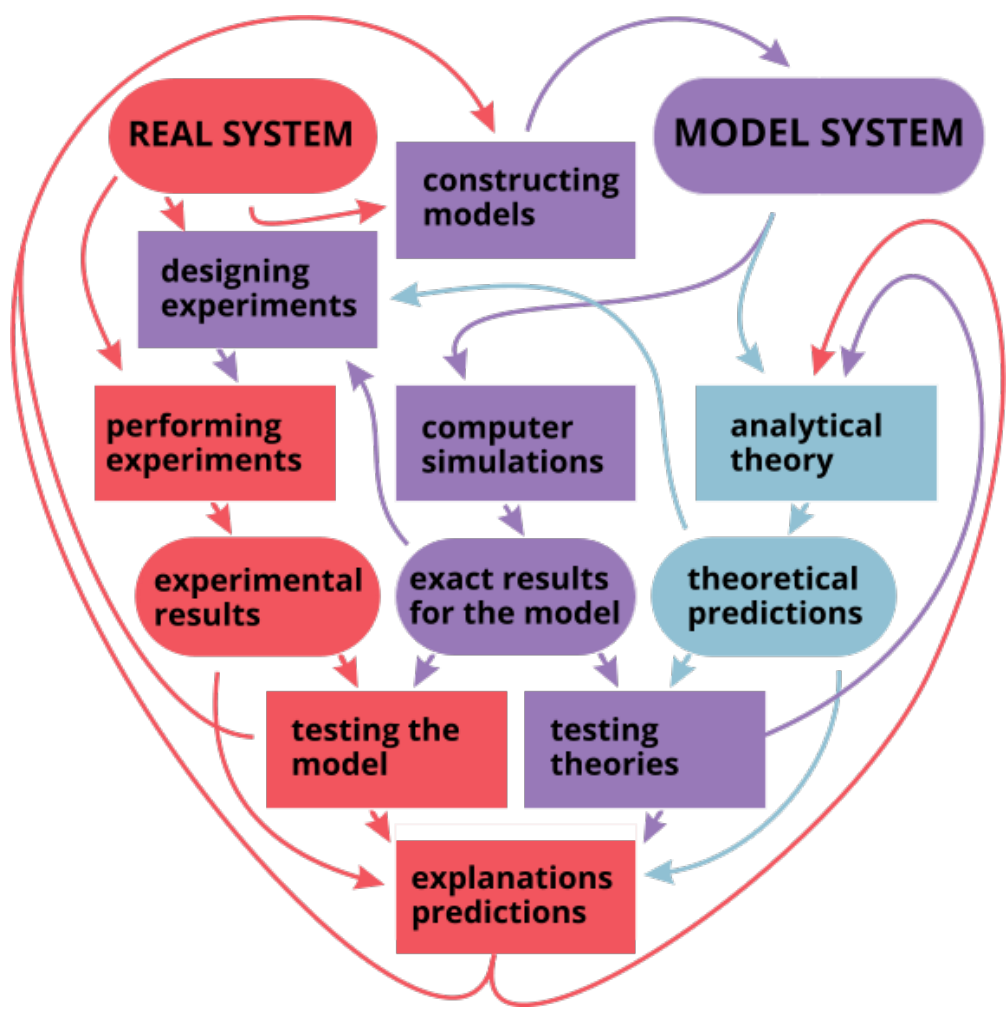

**Figure 1. Computer simulations are firmly established as the third paradigm of research alongside with experiments and simulations. Figure (slightly modified) by courtesy and permission from Markus Miettinen[15].**

The above does not mean that simulations are error free. As with all methods of research, there are errors, inherent, induced, accidental and due to ignorance. In this review, we will focus on practical issues in simulations rather than on a rigorous analysis of algorithms, numerical errors, errors due to finite precision and errors in data analysis. For integration algorithms, we refer to the book by Tuckerman[16]. He discusses a large number of integration methods and their derivations as well as constraint algorithms. Other recent discussions on integrators and algorithms can be found in Refs. [17-19]. For error analysis see, e.g., the article by Bond and Leimkuhler[20]. Sampling issues in MD have recently been reviewed by Grossfield and Zuckerman[21] and they are discussed in this issue in the article of Pomes et al.[22]. In addition to algorithms and protocols, system preparation should be paid due attention as has been pointed out by Manna et al.[23]. As a side note, in the context of numerical analysis, it has been pointed out that it is not clear at all why MD even works! [24]

Accidental errors can be loosely classified as programming errors and user errors including bad choices of simulations parameters. Even the best programmer makes about 15-50 errors per 1000 lines[25]. The advantage of open source software is, however, that errors are found very quickly. Indeed, the current authors have found and reported errors in open source software. The most dangerous category is errors due to ignorance, in other words, user error. Every error is ultimately a user error: The user has an obligation to check, verify and correct.

In the following, we discuss a few cases of various known errors. We hope this discussion encourages researchers to check software, parameters, protocols and, most importantly, their own practices. Using "the standard protocol" without having first-hand knowledge is very irresponsible. It is impossible to avoid all errors but the possibility of errors should always be minimized.

Before embarking on a more in-depth discussion, let us mention a couple well-known cases. One of the first ones that got major attention were the hidden errors in random number generation[26, 27]. This lead to re-evaluation of pseudo-random number generation algorithms and the development of more rigorous algorithms and testing protocols. Another famous case, is the subtle programming error that lead to a retraction of several protein structure prediction papers in the early 2000's[28]. This is significant because rather than blaming the programmer, one should always look at fundamental properties: In this case the protein structures predicted by the analysis code gave the wrong handedness (amino acids are left-handed, see e.g. Novotny and Kleywegt[29]). Despite this major issue, it took time to recognize the problem and the papers passed peer review in the so-called flagship journals and, most likely, influenced grant decisions concerning other researchers whose results did not agree with the faulty ones.

The rest of this paper is organized as follows: Next, we discuss two Case Studies based on our own research. After that, we discuss issues related to thermostats, Lorentz-Berthelot rules, electrostatics and some other matters in more detail.

## 2. Case study I: Unphysical flow of water inside a carbon nanotube

As our first Case Study, we discuss the behavior of water inside a carbon nanotube (CNT). The ability of water to enter the highly hydrophobic CNTs and the extremely rapid motions of water molecules inside a CNT were originally predicted by simulations[2] and later confirmed by both experiments[30, 31] and follow-up simulations[32-39].

Due to CNTs' high hydrophobicity and the consequent near frictionless transport of 'water, CNTs have emerged as promising candidates, e.g., for filtration applications[31, 33, 40]. As a consequence, many techniques to control the rate and direction of flow inside a CNT have been suggested[40-43]. Significant controversies, however, remain[32, 34, 40, 44-47]. This is what we will discuss next.

A pristine CNT is charge-neutral. Recently, MD simulations reported by Gong et al.[40] used charge-decorated CNTs to show that charge-induced ordering of water leads to spontaneous and continuous unidirectional flow through a carbon nanotube (CNT), Figure 2. They accredited this observation to the three discrete stationary electric charges placed asymmetrically along the exterior of the CNT. They chose this charge distribution such as to mimic the electric field inside an aquaporin protein[3] and proposed the observed spontaneous flow as a basis for designing novel spontaneously operating molecular pumps[40, 48].

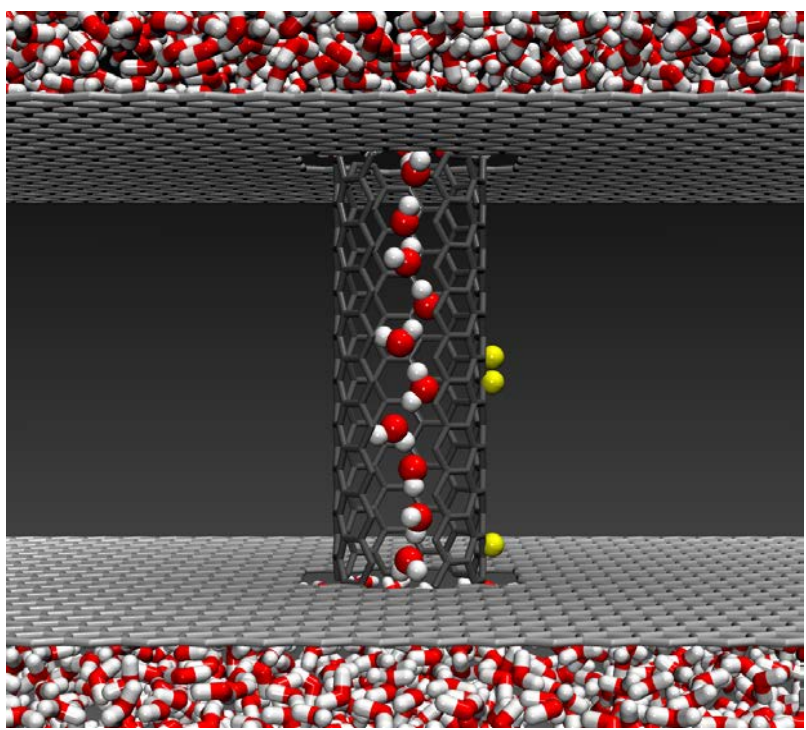

**Figure 2. A carbon nanotube connecting two water reservoirs (shown in red and white) with three static charges (yellow spheres) placed beside the nanotube to mimic the placement of charged amino acids inside aquaporin. The nanotube is 2.3 nm long and the charges of +e, +e/2 and +e/2 are placed 0.37 nm, 1.09 nm and 1.21 nm from the bottom of the nanotube, respectively. The overall system is charge neutral. Fully periodic boundary conditions.**

2.1. Three critical issues to prevent artificial flow:

Such a continuous flow without supplied energy is, however, a manifestation of perpetual motion and violation of the second law of thermodynamics. A simple gedanken experiment shows the impossibility of spontaneous unidirectional flux: If the observed flux was real, it would allow propelling a CNT in water without supplying any external energy[49]. A downward momentum of the flux would be compensated by an upward motion of the CNT. Gong et al. claimed that the energy required to produce the flow comes from the constraints imposed on the charges[40]. This is not the case[44]. Instead, the observed flow is a result of improper choice of simulation parameters and protocols: 1) Improper use of thermostatting, 2) careless application of a standard trick called neighbor list update frequency and, most of all, 3) incorrect use of charge groups[44]. In brief, incorrect use of charge groups caused a mismatch between the interactions of the three imposed charges and water molecules in the CNT. Moreover, other sources of simulation artifacts, improper use of thermostatting and neighbor list update frequency, can also lead to an artificial temperature gradient build-up within the CNT and thus cause an unphysical flow. With the physically correct flat temperature profile the artificial flow ceases. In addition to our studies[44], these findings have been confirmed by another independent study[45]. A more detailed discussion follows in later sections.

2.2 The effect of Lennard-Jones cutoff:

A net flux of water has also been observed in an uncharged CNT under an applied uniform electric field[50]. In this case, artifacts related to the treatment of the long-range part of the van der Waals interactions have been reported[51, 52]. In particular, Bonthuis et al.[45] simulated CNTs with different Lennard-Jones (LJ) cutoff lengths and two different truncation schemes: 1) simple cutoff, which sets the force to zero and the LJ potential to a small constant value beyond a cutoff distance ($r_c$). This results in a discontinuity at $r = r_c$. 2) A shifted cutoff, which makes the force to decay smoothly to zero by adding a nonlinear function[53]. The results of Bonthuis et al. show that with a simple cutoff the magnitude of water flux decreases as $r_c$ increases and approaches zero at the limit $r_c \rightarrow \infty$. With the shifted cutoff, the flux disappears for all $r_c$.

The above results cast doubt over the correctness of the implementation of the simple LJ cutoff in the GROMACS package[45, 51, 54]. Milicevic constructed a minimal model to examine the sensitivity of the simple cutoff scheme to the choice of $r_c$[52]. MD simulations of an uncharged spherical LJ particle in water, both in the presence and absence of an external electric field, were performed using GROMACS. A non-zero van der Waals force on the LJ particle under an external electric field was observed with the cutoff while the force vanished with the shift and switch treatment. The average forces were about −6.6 pN in simulations with the standard setup in all studied Gromacs versions of 3.3.3, 4.5.5, and 4.6.4. The authors concluded the software does not have errors and the observed artifacts are due to the cutoff. This study further suggested that the non-zero force was generated by water molecules belonging to the first two hydration shells. Although, the result is sufficient to conclude that a non-zero force is produced by simple truncation of vdW forces, understanding of water's first two hydration shells causing the non-zero forces has not been achieved [52].

One can also speculate that such effects might be due to integrator algorithms. Although we are not aware of systematic studies addressing the issue in this context, Toxvaerd and Dyre[55] studied cutoff and force shifting in Lennard-Jones systems and concluded that shifting the forces smoothly to zero is important. This is in line with Bonthuis et al.[45, 51] In addition, in a follow-up paper Toxvaerd used the smoothly shifted forces in his studies of stability of MD simulations and came to the conclusion that with normal time steps, the standard Verlet algorithm is indeed excellent.

## 3. Case study II: Unphysical flow inside a protein nanochannel

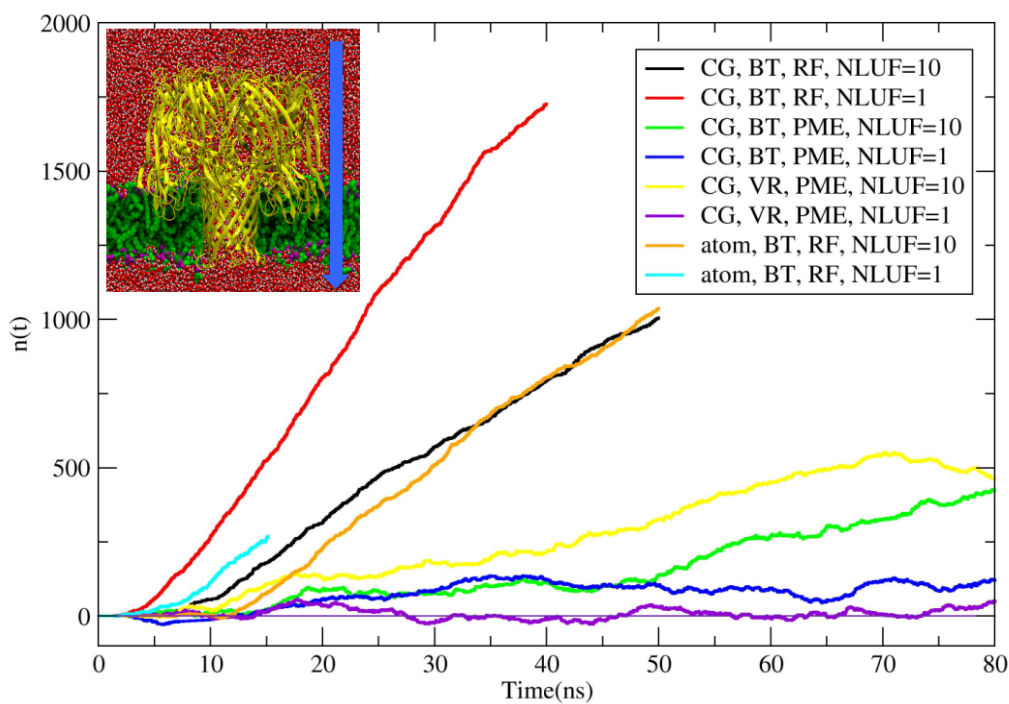

**Figure 3. Net flux of water molecules ($n(t)$) permeating from the cap to the stem as a function of time. The blue arrow indicates the direction of the net flow. Simulation parameters for each system are provided in the legend. Charge group (CG) and atomic group (atom)[53] were used as a basis to compute the neighboring cut-off. For charge groups, the default values of the Gromos 53A5 force field[56] were used. Thermostats: Berendsen weak coupling thermostat (BT)[57] and the velocity-rescale (VR) thermostat [58, 59]. Long-range electrostatic interactions: reaction field (RF)[60] and Particle Mesh Ewald (PME)[61, 62]. The neighbor list update frequency (NLUF) was varied between 1 and 10 time steps[53]. Inset: A snapshot from a simulation. Protein (yellow ribbon) is embedded in a palmitoyloleoylphosphatidylcholine (POPC) lipid bilayer (green spheres). The system was hydrated by simple point charge (SPC) water molecules[63] (red spheres) with a density of about 1000 kg/m³. Periodic boundary conditions were applied.**

Artificial unidirectional flow also appears in membrane embedded protein systems when the simulation parameters are not chosen carefully[64]. Due to the complexity and inhomogeneity of the system, the combinations of simulation parameters such as the thermostat algorithm, treatment of electrostatic interactions, charge groups and neighbor list update frequency must all be carefully considered and tested[64] as in the case of CNTs. Figure 3 shows that an artificial net flux appears very easily unless *all* the relevant simulation parameters are chosen properly. It is also important to notice that none of the parameter combinations shown in the Figure 3 would produce artifacts in a pure membrane system. The introduction of a membrane embedded protein creates a very different (inhomgeneous) local environment.

Next we discuss the parameter choices in more detail. As Figure 3 shows, the system appears to be particularly sensitive to the choice of the neighbor list update frequency (NLUF in the legend). Updating the neighbor lists is very time consuming, and thus less frequent updates are preferred for computational efficiency. Update frequency of 10 (every $10^{th}$ time step) is the most commonly used choice in membrane simulations and also the default value in Gromacs. This technique is commonly used since it provides a significant speed up in the calculation of forces. However, some interactions could be missing from the calculation if the lists are not updated often enough. The situation is, however, somewhat more subtle. As Figure 3 shows, the net flux with NLUF=1 may be even larger than with NLUF=10 depending on other parameters. This demonstrates an unexpected and potentially dangerous cancellation of errors.

In addition to the neighbor list update frequency, the method of computing the electrostatic interactions has a major effect. As Figure 3 shows, simulations using the Particle Mash Ewald (PME) technique[61, 62] perform significantly better in comparison to the reaction field (RF) method (simple cutoff should *never* be used). Qualitatively, this difference can be traced to the fact that the dielectric constant of the system is not homogenous because of the molecular mixture of protein, lipids and water. PME takes these issues into account much more accurately than the reaction field which uses a mean-field beyond the immediate vicinity of each charge. In addition, at the RF cutoff, there is a discontinuity in the electrostatic potential and this may create an artificial force[55, 65]. Recently, Ni et al.[65, 66] reported that the use of a charge group based truncation with the reaction field can lead to an artificial repulsion between charged residues. We tested this by comparing simulations using NLUF=1 and NLUF=10 and atom based truncation. The results in Figure 3 show that the unidirectional flow persists.

The one remaining issue is thermostatting. In our previous study[44], we have shown that the Berendsen weak coupling thermostat could create an inhomogenous temperature inside a nanopore leading to a net flow. When the velocity-rescale algorithm[58, 59] was applied instead, the unidirectional artificial flow disappeared.

The effect is also visible in Figure 3, but it is weak compared to the other two effects discussed above. In summary, a safe protocol for simulations of systems containing pores and channels to avoid an artificial unidirectional flow should contain: i) PME (or comparable) for the treatment of long-range electrostatic interactions, ii) neighbor lists should be updated at every time step, and iii) the velocity-rescale thermostat instead of the Berendsen weak coupling thermostat[64].

## 4. Long-range electrostatics

We start the more general discussion with electrostatic interactions. This topic was already mentioned in the two Case Studies above and has been discussed in numerous reviews, see e.g., see Refs. [67-71]. In general, since the computation of electrostatic interactions is time consuming, several methods have been introduced to deal with them. Earlier methods include truncation of the Coulomb interaction. It has been shown in a number of studies that it may lead to errors and severe artifacts. In an earlier paper, Saito showed that protein simulations are strongly influenced by truncation errors[72]. On membrane systems, one of us showed that evaluation of the long-range electrostatic interactions by cutoffs introduced spurious effects and even lead to a phase change in lipid bilayers[73, 74]. Currently, virtually all modern-day simulations are performed using Ewald summations based methods such the Particle-Mesh Ewald (PME)[62] and Particle-Particle-Particle-Mesh (P3M)[75, 76]; methods such as the Fast Multipole Method[77, 78] and multigrid[79] would work too but they have yet to gain more popularity.

The Ewald summation and other Ewald based methods have also been questioned. An earlier study suggested that there are finite-size artifacts related to the Ewald method which have their origin in the difficulty of defining the "zero of energy"[80]. The possible artificial periodicity and the resulting rotational properties were further studied by Smith and Pettitt[81] who concluded that for liquids with high dielectric constant (aqueous solutions) artifacts are negligible.

Another issue related to electrostatics is boundary conditions as discussed in the above reviews. In general, the straightforward implementation of any of the Ewald methods requires periodic boundaries to be efficient, yet in many cases it might be interesting to simulate systems that do not have fully periodic boundaries. We will not discuss this in detail but refer the reader to Refs. [68, 82] In addition, there has been recent interest in simulating systems with net charge. The Ewald based methods require charge neutrality. Extensions to non-charge neutral systems have been suggested and artifacts related to them have been recently discussed by Hub et al.[83].

Charge groups are used to increase performance for updating the Coulomb interactions of neighboring atoms and they can provide a remarkable speedup[53, 67]. Although the electrostatic interactions are calculated for all individual charges in the group, the locations of charge groups are defined by the averages of the coordinates of the charges belonging to the group, see Refs. [53, 67] for details. Unfortunately, it is very easy to reproduce the unphysical flow by using the default simulation parameters.

Computational cost has been one of the main concerns with systems rich in charges (partial and otherwise) such as biomolecular systems. In the past, the most common technique was to use truncation instead of the full Ewald summation. That offers a considerable saving in time since truncation offers $O(N)$ scaling whereas the plain Ewald summation scales as $O(N^2)$ or after some tricks as $O(N^{3/2})$. The particle-mesh approaches, Particle-Mesh-Ewald (PME) and Particle-Particle-Particle-Mesh (P3M), scale as $O(N \log N)$. There is a catch, though: As already mentioned above, several studies have shown that using truncation can lead to spurious effects and may have a strong influence on the physical and dynamical properties of biological systems[72, 73, 84-89]. There are several recent reviews discussing the treatment of electrostatic in soft matter systems[67, 68, 90, 91].

## 4.1 Electrostatics and protein folding simulations

In the area of protein folding simulations, the results can be sensitive to the choice of a specific scheme. For example, effective melting of a double-norleucine mutant of villin[92] has been observed at 380 K in simulations employing the PME method for the calculation of the long-range electrostatics and a 9.0 Å cutoff for van der Waals forces[93]. Meanwhile, a melting temperature of 300 K was reported when the reaction field method was employed for electrostatics with a cutoff of 8.0 Å[94, 95]. Piana et al.[96] investigated how the treatment of electrostatic interactions affects the free energy of folding and the structural properties of proteins[97]. The simulations were performed with two different schemes, a cutoff-based force-shifting technique[98] and PME[99] for the treatment of electrostatic interactions. The length of the atom-based cutoff was varied ranging between 8.0 and 12.0 Å to determine both electrostatic and van der Waals interactions. The simulation results showed that the free energy of folding of a small protein is insensitive to the choice of approximation methods when a cutoff beyond 9 Å is used but structural properties of the unfolded state depend more strongly on the approximation scheme and parameters[96].

## 4.2 Artifacts due to atom- and group-based truncations with reaction field electrostatics

The treatment of non-bonded interactions is an extremely important issue in MD simulations[100, 101]. As already discussed, one of the most common and simplest approaches is truncation but the disadvantage is computational inaccuracy with possible severe artifacts [102, 103]. The LJ potential is considered to be of short-range; generally, interactions are considered to be long-range if the exponent α in $1/r^\alpha$, satisfies $\alpha < d$ where $r$ is the interparticle distance and $d$ is the dimension of space.

With periodic boundary conditions, the PME method is widely used since it is more accurate than the truncation schemes, although it may be computationally more expensive than truncation. PME could, in principle, be also used for computing LJ interactions. Whether or not truncation is faster than PME depends, however, on the cutoff distance. Although generally not much of concern, some undesirable artifacts due to the interactions between the central cell and its periodic images have been reported[104, 105]. An alternative approach is based on continuum electrostatic theory, in which the continuum part is evaluated through a reaction field (RF)[60, 106, 107]. Since only pairwise interactions are accounted for, the RF method is reasonably fast in comparison with PME. However, as already mentioned above, one the major disadvantages of the RF approach is that it was developed specifically for homogeneous systems[60, 106, 108]. In practice, several recent studies have used the RF-method in the simulations of ions[109], DNA/RNA/short peptide molecules in aqueous solutions[110-115] and lipid bilayers[84]. Some of these simulations[103, 110] reported satisfactory results with the RF method while others[84, 116] did not.

Recently, Baumketner [65] found that the success of the reaction-field method applied to ions critically depends on the implementation of truncation between the solute and the solvent. Truncation using charge groups was reported to cause artificial repulsion between charged solutes, while using an atomic basis (each atom is a separate group) gave good agreement with PME [65]. Ni et al.[66] also studied how group- and atom-based truncations influence reaction-field simulations of proteins and DNA. Truncation based on charge groups was observed to disrupt native states even in short nanosecond simulations due to artificial repulsions between charged groups forcing the molecules to unfold. Atom-based truncation produced stable trajectories for both DNA and proteins, and the results were in good agreement with experiments and simulations using PME. Even though there are systems for which the group truncation is adequate[103, 110], it is difficult to formulate a quantitative criterion to avoid these artifacts. The decision whether to use group-based or atom-based truncation should be carefully made on a case-by-case basis.

## 5. The need to go beyond the Lorentz-Berthelot rules

For simulations of mixtures of different atoms, virtually every book on molecular simulation introduces the Lorentz-Berthelot rules of mixing (see, e.g. [14]): When simulating a mixture of atoms of different types, the Lorentz-Berthelot rules state that the Lennard-Jones parameters, σ and ε can be obtained as simple arithmetic and geometric averages, respectively, as

$$\sigma_{ij} = \frac{\sigma_{ii} + \sigma_{jj}}{2}$$

$$\varepsilon_{ij} = \sqrt{\varepsilon_{ii}\varepsilon_{jj}},$$

where the indices $i$ and $j$ denote the different atom types, and $\sigma_{ii}$ and $\varepsilon_{ii}$ for all $i$ are known. The first part was introduced by Lorentz[117] in his 1881 article and the second part, the geometric mean for the depth of the energy well, is due to Berthelot[118]. The Berthelot rule is sometimes justified by using London's dispersion theory (see e.g., Ref. [117]) and approximations for the ionization potentials and molecular sizes. See Rowlinson and Swinton[119] for a detailed discussion of the various aspects of the Lorentz-Berthelot mixing rules. Although a derivation along the lines stated above can be provided, the Lorentz-Berthelot rules are essentially a pragmatic choice that works reasonably well in many cases.

The Lorentz-Berthelot rules are not the only possible choice, but they remain the most common one and are used, for example, in the CHARMM[120] and Amber[121] force fields. It is also possible to construct other combining rules based on the same theoretical approach but using different approximations with regard to ionization potentials, molecular sizes and not using the hard sphere approximation. For example, OPLS[122] uses the geometric mean for both σ and ε as does the GROMOS[123-125] force field. Other combination rules, although based on a similar approach, are the more specific ones such as Waldman-Hagler[126], Kong[127], Halgren[128] and Fender-Halsey[129] rules. Since the Lorentz-Berthelot are used in some of the most popular force fields, it is clear that these specific rules are not as commonly used. This may change in the future.

There has been growing interest in going beyond Lorentz-Berthelot. This is due to the increasing number of explicit demonstrations in which the Lorentz-Berthelot rules fail in a rather spectacular manner[130-134]: For example, Delhommelle and Millié[130]showed that the application of the Lorentz-Berthelot rules may lead to incorrect thermodynamic properties for simple binary mixtures whereas using the Kong rules[127] yields good agreement with experiments. In another study, one of us demonstrated that in simulations of atoms in the gas phase interacting with a surface, the Lorentz-Berthelot rules give surface-gas interactions that are 10 times stronger than they should be[133, 135]. This study also showed that the standard Lennard-Jones potential may be too hard; the softer Morse potential[14] provided a much better fit to data from *ab initio* simulations. Chase et al. also found similar problems in obtaining the proper energy well depths[136].

From the above and other studies it has become evident that in addition to parameterization against experiments and quantum-chemical simulations, attention needs to be paid to the mixing rules / parametrization of unlike-atom interactions. As a very recent example, Nikitin et al.[137] proposed a new Amber-ii compatible force field for perfluoroalkanes not using the usual Lorentz-Berthelot rules. This new force field demonstrated good agreement with both experiments and quantum level calculations. The authors point out, however, that the new parameterization is no longer compatible with the common Amber/OPLS force field. Other works along these lines include Vlcek et al.[138], Hu and Jiang[139], Duarte et al.[140], Forsman and Woodward[132], and Moucka et al.[141], Rouha and Nezbeda[142]. Problems with the standard forms were also noticed in water-protein interactions by Piana et al.[143]. We expect that similar new parameterizations will be introduced for other types of molecules as well.

**6. Force fields: A very brief summary**

We will not discuss force fields here. Force fields are discussed in this issue by Lyubartsev et al.[144] and for completeness, we also refer to some additional recent reviews [145, 146], tests [93, 147-149] and developments[150].

**7. Thermostats – necessary evil**

MD simulations in their basic formalism are inherently in the NVE (constant particle number, volume and energy) ensemble. To simulate the experimentally relevant canonical (NVT) ensemble, a thermostat must be added to maintain constant temperature. The addition of a thermostat may significantly affect the thermal fluctuations in the system and cause energy drifts that sometimes have their origin in accumulation of numerical errors[151-155]. This is particularly the case when truncation schemes are used for interactions: Conservation of linear and angular momenta may be violated and although these quantities are initially set to zero, their values will unavoidably change due to numerical errors.

In addition, thermostats can be roughly divided into *local* and global. Global thermostats apply a uniform change instantaneously to all the atoms of the system. Nosé-Hoover[156, 157] and the Berendsen weak coupling method[57] are examples of them. Local thermostats, on the other hand, act on individual atoms or pairs and typically employ fluctuation-dissipation relation from statistical mechanics. We will discuss both of these categories below.

7.1 The "flying ice cube" effect and what to do about it

One of the best known artifacts in MD simulations is the so-called "flying ice cube" discussed first by Harvey et al.[151]. The "flying ice cube" artifact gets its name from the fact that when a system suffers from it, its high-frequency motions drain to low-frequency modes and eventually the system freezes and becomes a flying ice cube; the center of mass motions, both rotation and translation, are low-frequency motions and gain energy due to the drain from high-frequency motions. In their original paper, Harvey et al. pointed out that the flying ice cube artifact is a violation of the equipartition principle and is not specific to any simulation package but rather solely due to velocity rescaling – a technique applied by global thermostats including the commonly used Berendsen weak coupling method[57] and Nosé-Hoover[156, 157]. This problem is typically associated with the Berendsen weak-coupling method but Wagner et al.[158] have shown that it also occurs in constrained internal coordinate simulations with the Nosé-Hoover thermostat with very large (20 fs) time steps. Harvey et al.[151] suggested three approaches to remedy the problem: The first one is velocity *reassignment* instead of *rescaling*. This essentially means using a local thermostat that acts on individual atoms or pairs, rather than on all the atoms with the same strength. Andersen[159], Lowe-Andersen[160], Langevin[161] and the dissipative particle dynamics thermostat[162] are examples of such; commonly used terminology for dissipative particle dynamics often refers to a soft conservative potential together with a fluctuation-dissipation relation, but the method is independent of the precise form of the conservative potential and can be considered as a momentum conserving Langevin thermostat[163]. These local thermostats apply Gaussian random noise to maintain the canonical ensemble. The other two suggestions were removal of center of mass motion and better algorithms for rescaling. The latter two are typically part of modern simulation protocols: In a typical simulation, center of mass motion is removed periodically during the simulation and on the algorithmic side the practical remedy is to use sufficiently large coupling times for global (such as Berendsen weak coupling) thermostats. A modification of the Berendsen weak coupling method, the so-called velocity rescale algorithm of Bussi, Donadio and Parrinello[58, 59] has been shown[64] to perform well and has lately become popular.

An important question related to thermostats is sampling, i.e., how to determine if the ensemble is reproduced and sampled adequately. This is crucial for computing thermodynamic properties. An elegant procedure for testing this based on performing paired simulations and cancellation of system-dependent properties between them was recently introduced by Shirts[164]. He tested several thermostats and barostats and concluded that of the thermostats, "all tested thermostats except the Berendsen thermostat give statistically good results". The tested thermostats included the Berendsen weak coupling[57], Nosé-Hoover[156, 157], the stochastic Langevin[161], Andersen[159] and the velocity rescale algorithm of Bussi, Donadio and Parrinello[58, 59]. We will discuss the results for barostats below in Sec. 8. Shirts also provides tools and code as open source at http://simtk.org/home/checkensemble. For further reading on different aspects of thermostatting, please see also Refs. [165, 166]. Finally, we would like to point out that in the case of non-equilibrium systems, thermostatting becomes even more complex[167].

## 7.2 Failure of Nosé-Hoover for non-ergodic systems and the fix

A lesser known, although well-documented issue is the failure of the Nosé - Hoover thermostat for non-ergodic systems. Integrable systems are examples of such. In large ergodic systems – such as virtually all simulations of biological systems - no such problem exists. The solution was given by Martyna, Klein and Tuckerman[168] who connected a series of Nosé-Hoover thermostat to construct a Nosé-Hoover chain. In current protocols, Nosé-Hoover chains are commonly used in *ab initio* simulations. An excellent discussion of both the Nosé-Hoover method and Nosé-Hoover chains is provided in the book of Tuckerman which also discusses methods for Hamiltonian systems including the Nosé-Poincare method of Bond et al.[169].

## 7.3 Hot solvent – cold solute problem

In simulations of biological systems, such as membranes, it is common to apply separate thermostats to the solvent (water and ions) and the solute (e.g. a membrane). This is done to avoid stationary temperature gradients due to too slow exchange of kinetic energy in inhomogeneous solvent-solute systems: With a single (global) thermostat, the solvent and solute may acquire different temperatures. This is the so-called "hot solvent – cold solute problem"[170]. The simplest solution is indeed to apply thermostats separately the solute and solvent. In practice, the simulated system may consist of macromolecules and water, and a separate thermostat can perturb the dynamics of the macromolecule much more strongly than single global control. This may introduce large artifacts into the conformational dynamics of the macromolecules. Recently, Lingenheil et al.[171] proposed a strategy of controlling temperature based on the concept that an explicit solvent environment represents an ideal thermostat concerning the magnitude and time correlations of temperature fluctuations of the solute. This strategy can provide a homogeneous temperature distribution with the correct statistical ensemble and minimal perturbation dynamics of the solute molecule.

## 7.4 The effect of treatment of thermostat on protein folding in the replica exchange molecular dynamics (REMD) simulations

Not only can the approximations of long-range interactions produce a significant problem in computations of thermodynamic, structural, and dynamic properties of proteins but a thermostat can also have an effect on the folded/unfolded state populations[153, 172]. The problem when using the Berendsen weak coupling thermostat in REMD simulations was pointed out by Rosta and Hummer[153, 173, 174]. REMD is a method to enhance conformational sampling in MD simulations by using several copies of the physical system in parallel at different temperatures[175, 176]. Then, attempts to exchange the structures between these systems are made at certain intervals to increase the efficiency of conformational sampling. The acceptance criterion to exchange conformation between any two temperatures is designed to maintain the canonical probability distribution in the configurational space. As already mentioned above, it has long been known that the Berendsen weak coupling thermostat does not correctly reproduce the canonical ensemble[154, 155]. Rosta and Hummer investigated its effects on protein folding[153] and observed a shift in the equilibrium folded/unfolded states of a small peptide in water. They showed this to be related to the weak coupling thermostat. The folded state was observed to be overpopulated and underpopulated at low and high temperatures, respectively. The population shift results from the narrowed down potential energy distribution due to the non-canonical ensemble[153]. Thus, REMD simulations of proteins folding should only be performed with thermostats that produce the canonical ensemble such as the Langevin thermostat[161, 177], Andersen thermostat[159], Nosé-Hoover thermostat[156, 157] or the velocity-rescale thermostat[58, 59].

## 8. Barostats: How about compressibility?

Simulations using the NpT (constant particle number, pressure and temperature) ensemble are typically needed in simulations of membrane systems. To simulate the NpT ensemble, a barostat is needed in addition to a thermostat. Common choices include the Langevin piston[178], the Rahman-Parrinello method[179], the

Martyna-Tuckerman-Tobias-Klein algorithm [180, 181] and, perhaps most commonly, the Berendsen barostat[57]. The Berendsen method is conceptually simple and hence it is very commonly used. It is often used in membrane simulations to evaluate compressibility but, strictly speaking, that is not entirely correct since this barostat does not produce the correct canonical distribution[182]. This is important since the area compressibility is typically evaluated using the formula $K_A = A\left(\frac{\partial \gamma}{\partial A}\right)$, where $K_A$ is area compressibility, $A$ the area, and $\gamma$ surface tension. The Parrinello-Rahman barostat, although computationally more expensive, produces the correct distribution and does not suffer from such issues. Please see Waheed and Edholm[183] for an in-depth discussion about compressibility of bilayer systems.

As discussed in Sec. 7.1, Shirts[164] also tested barostats. In his tests, the Martyna-Tuckerman-Tobias-Klein barostat performed the best and the Parrinnello-Rahman was determined to be "acceptable". For the commonly used Berendsen barostat Shirts writes: "Berendsen pressure control is simply wrong for any calculation where volume fluctuations are important." In the context of membrane (and comparable) simulations using the NpT ensemble, this means that elastic moduli, such as compressibility, will not be correct. In addition, Shirts noticed abnormally long autocorrelation times when the Berendsen barostat was used together with the Bussi-Donadio-Parrinello velocity rescale thermostat. The reason remains unresolved. We would like to note that Shirts' tests were carried out using a very long 8 fs time step (typical MD time step is 2 fs). Shirts used Gromacs 4.6 for the simulations. In another study, Rogge et al.[184] used metal-organic frameworks and came to similar conclusion: The Martyna-Tuckerman-Tobias-Klein and Langevin barostats were shown to work well in such systems. The Rahman-Parrinello method was not tested.

## 9. Epilogue: User - the most significant error source

The above discussion listed some of the common sources of error in molecular simulations. We also presented two case studies to show some of the effects of such artifacts and how unexpected yet potentially dangerous cancellation of error may occur. It is important to notice that some of the topics discussed are not actual artifacts. For example, the trick of updating the neighbor lists only about every 10 time steps is well validated and works extremely well in many cases. When used improperly it will not (unless used really recklessly) make a simulation crash.

Improper use will, however, yield unphysical results as discussed. One may ask, whose fault is it? The answer is simple: It is *always* the user's fault. No matter how simple the simulation, the user must always *check* and *validate* all the parameter (as well as protocol) choices even if they have been used extensively before. Breaking the second law of thermodynamics like in Case Studies 1 and 2 demonstrates that anything is indeed possible and that such unphysical results may look very exciting. As with many other things in life, if something looks too good, it most likely is. This also explains why so many bad, uninformed, or sometimes old, choices still remain in simulation protocols: Validation is time consuming and thankless. The best case is that everything is ok. That is in some sense frustrating as it appears to be a waste of time. In the worst case, something appears to be wrong or suspicious and finding the origin may be extremely time consuming and tedious as anyone who has tried it knows very well. Sometimes there is a silver lining: A new protocol or method is established and published, serving the whole community.

It may sound absurd and somewhat provoking, but available simulation software (i.e., not home grown programs) becoming very user friendly and easy to obtain can almost be seen as a negative development! It is very easy for anyone to download GROMACS [185-187], NAMD[188], LAMMPS[189] or any other powerful simulation package and set up simulation without having the faintest idea of the algorithms, protocols, tricks of the trade, analysis, etc. Since all of those packages represent the state-of-the-art, they are both very powerful and algorithmically stable – two features that may give an uninformed user the false feeling of power and knowledge: One does not become a theorist by buying chalk, experimentalist by buying a microscope or a computational scientist by downloading software. Chalk may be the most challenging for getting the initial results out, but with a microscope and software one gets, if nothing else, pretty pictures very easily. Pretty pictures are not results of rigorous research. In able hands, however, computational modeling is an extremely powerful tool that provides quantitative results, has true predictive power, and can steer research to a new track.

Although we have discussed errors, artifacts and problems, we would like to emphasize that this is a sign of an active and healthy field – identification of problems leads to improvements, higher reliability, quantitative predictions and new developments. Computational research is at a very active and exciting state and we are positive that this trend will continue for a long time.

Finally, here are some recommendations that should serve as good starting points for basic biomolecular simulations based on current knowledge: The Bussi-Donadio-Parrinello velocity-rescale thermostat[58, 59], the Martyna-Tuckerman-Tobias-Klein[180, 181] and Langevin barostats[161]. If full Ewald-based summation is not used for Lennard-Jones, then forces should be shifted smoothly to zero. PME[61, 62] or P3M[190, 191] for long-range electrostatics (for the cases with non-periodic boundary conditions, please see the discussions in Refs.[67, 68]). And always remember to test!

## Acknowledgements

J.W. gratefully acknowledges financial supports from the Faculty of Science and Kasetsart University Research and Development Institute (KURDI) at Kasetsart University.